\documentclass{bioinfo_}
\copyrightyear{2013}
\pubyear{2013}

\usepackage[T1]{fontenc}
\usepackage{graphicx}
\usepackage{setspace}

\usepackage{color} 
\usepackage{framed} 
\usepackage{soul}
\usepackage{xcolor}
\colorlet{shadecolor}{yellow}
\newcommand{\hilight}[1]{#1}

\usepackage{amssymb,amsfonts,amsmath}
\begin{document}

\firstpage{1}

\title[Reconstructing protein binding patterns from ChIP time-series]{Reconstructing protein binding patterns from ChIP time-series}
\author[Sch\"olling \textit{et~al}]{Manuel Sch\"olling\,$^{1}$, Rudolf Hanel\,$^{1,}$
\footnote{to whom correspondence should be addressed}}
\address{$^{1}$Section for Science of Complex Systems/CeMSIIS, Medical University of Vienna, A-1090 Vienna, Austria}

\history{}

\editor{}

\maketitle

\begin{abstract}

\section{Motivation:}
Gene transcription requires the orchestrated binding of various proteins
to the promoter of a gene. The binding times and binding order of
proteins allow to draw conclusions about the proteins' exact function
in the recruitment process. Time-resolved ChIP experiments are being used to analyze
the order of protein binding for these processes. However, these ChIP signals
do not represent the exact protein binding patterns. 

\section{Results:}
We show 
that for promoter complexes that follow sequential recruitment dynamics the ChIP
signal can be understood as a convoluted signal and propose
the application of deconvolution methods to recover the protein binding
patterns from experimental ChIP time-series. We analyze the suitability of four deconvolution methods: two non-blind
deconvolution methods, Wiener deconvolution and Lucy-Richardson
deconvolution, and two blind deconvolution methods, blind Lucy-Richardson
deconvolution and binary blind deconvolution. We apply these methods
to infer the protein binding pattern
from ChIP time-series for the \textit{pS2} gene.

\section{Contact:} \href{rudolf.hanel@meduniwien.ac.at}{rudolf.hanel@meduniwien.ac.at}


\end{abstract}

An essential step in gene expression is the initiation of transcription.
This process requires the orchestrated recruitment of regulatory
proteins to the gene promoter. This leads to the formation of the transcriptional
machinery that transcribes a gene from DNA to RNA. Transcription
factors assemble on the promoter site, forming sequences of 
protein complexes on the promoter. Eventually, the protein complex
attracts the RNA polymerase that transcribes the gene and finally
the promoter is cleared again. During this
process the modification of epigenetic marks on the DNA and histones
were shown to be essential for the initiation of transcription \citep{Metivier2006}. 

For various promoters the proteins participating in the complex formation have been identified,
yet, determining the exact order and timing of the recruitment events is still a non-trivial task.

Extensive ChIP experiments have been used to examine the binding order
of proteins and the modification of epigentic marks \citep{Metivier2006,Lee2005}.
Since a large number of cells ($\approx 10^6$) are required to perform ChIP experiments \citep{Metivier2006},
such experiments are performed on initially synchronized cell populations
with a cleared promoter site. 

Usually ChIP experiments are analyzed heuristically by applying prior knowledge of protein interactions \hilight{to interpret the form
of the ChIP signal.}

In \cite{Pigolotti2007} the sequence of maxima and minima in ChIP time-series was 
used to predict \hilight{the structure of the dominant negative feedback loop
that governs the recruitment dynamics.}

In \cite{Hanel2012} ChIP data was analyzed by representing the recruitment
process as a regulatory network. The proteins involved in
the recruitment process are represented by nodes in this network.
By linearizing the regulatory dynamics, one can determine how the
binding of a protein affects the affinity of other proteins to bind.

By assuming a circular recruitment process that is traversed stochastically in each cell, \cite{Lemaire2006a} reproduced
the binding pattern by a least-square fit of a ChIP signal against simulation data of
the recruitment process. This method results in binding patterns that show \hilight{multiple} binding times for most proteins 
for the $pS2$ promoter. These binding patterns have been interpreted as stochastic binding events that are seen for example
with other experimental methods like GFP (\textit{green fluorescent proteins}).

\begin{figure}[!tpb]
\begin{center}
\includegraphics[width=86mm,height=4cm,keepaspectratio]{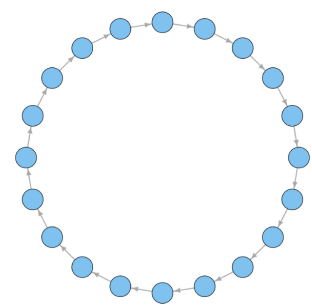}
\end{center}

\caption{Visualization of the recuritment network for sequential recrutiment. The recruitment matrix $a_{ji}$ circularly links the recrutiment states and each state has only one successor state.
\label{fig:seqrec}
}

\end{figure}

In \cite{Scholling2013} we generalized this recruitment model to address the question of whether the
binding order in a recruitment process is deterministic (sequential recruitment) or stochastic
(probabilistic recruitment).
In this model the recruitment process is represented by a walk on a network of recuritment states.
In the case of sequential recruitment each recruitment state has only one possible successor state (compare Fig. \ref{fig:seqrec}) whereas
in the case of probabilistic recruitment the topology of the network can look much more complex.
We showed that in a cell population the occupation $p_j(t)$ of a recruitment state $j$ at time $t$
is given by
\begin{equation}
  \label{prob}
  \frac{d}{dt} p_j(t) = \sum_{k=1}^N \left(a_{jk} - a_k \delta_{j,k}\right) p_k(t)
\end{equation}
where $a_{jk}$ characterizes the transition rate from the recuritment states $k$ to state $j$, 
$\delta_{j,k}$ is the Kronecker symbol and $a_k=\sum_i a_{ik}$. If $a_{jk} = 0$, a transition
from state $k$ to $j$ is impossible.
Following \cite{Scholling2013} the resulting ChIP signal $c_m(t)$ for a protein $m$ in a
circular sequential recruitment process with frequency $\mu$ can be predicted by
\begin{equation}
\label{convolution}
c_m(t) = \int_0^1 S_m(x-\mu t) P_0(x) dx
\end{equation}
in the limit of a large number of recruitment states
($N \rightarrow \infty$), where the periodic function $S_m(x)=S_m(x+1)$ denotes
the protein binding pattern of a protein ($S_m(x)=1$ if protein $m$ is bound in
state $x$ and $0$ otherwise), $x\equiv k/N$ and $P_0(x) \equiv p_{xN}(t=0)$.
\hilight{
The assumption of large N can be justified in the context of the pS2 gene}
\citep{Scholling2013,Metivier2003}.

Consequently, the ChIP signal $c_m(t)$ does not directly represent the binding pattern
$S$ but it is ``blurred'' by $P_0$.

Based on this observation, we analyze methods to recover the protein binding pattern
from ChIP time-series by deconvolution in this paper. Deconvolution
methods attempt to recover the original, un-blurred source signal from
a measured blurred signal. 
The problem of deconvolving signals is found in various fields of research, ranging
from image processing \citep{Fergus2006,Schulz1997,Cannon1976} over engineering \citep{McDonald2012}
and spectroscopy \citep{Nadler1989} to biology \citep{Down2008,Lun2009}.
In the context of ChIP time-series, we assume that the synchronization of 
a cell population is imperfect, e.g. due to variation in the time required by
each cell to translocate proteins into the nucleus \citep{Ashall2009}.

Deconvolution can be performed using various mathematical approaches,
which are based on different assumptions on the source signal and/or
the blurring process. In this paper we discuss the
application of four deconvolution methods to recover the protein binding
pattern from ChIP time-series: Wiener deconvolution, Lucy-Richardson
deconvolution, blind Lucy-Richardon deconvolution and binary blind
deconvolution.

We demonstrate the application of these four deconvolution
methods on ChIP time-series data for the \textit{pS2} gene \citep{Metivier2003} and discuss
the results of the different methods.

\begin{figure}[!htpb]
\begin{center}
\includegraphics[width=86mm,height=9cm,keepaspectratio]{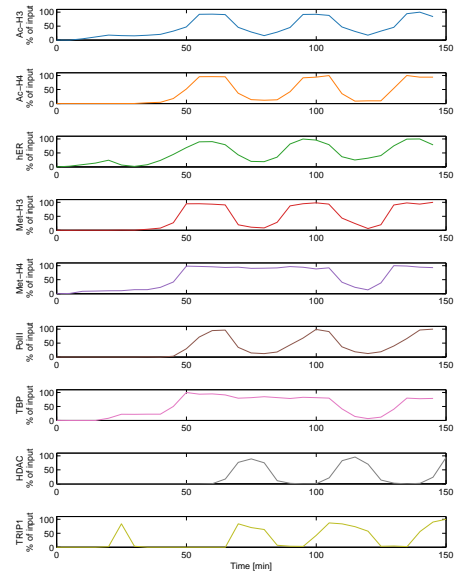}
\end{center}

\caption{ChIP time-series for the pS2 promoter:
The time-series show the ChIP signals (normalized) for five proteins and modifications
(methylation and acetylation) of two histones. Each time-series covers
the first, unproductive cycle and three productive cycles
(data extracted from \cite{Metivier2003,Lemaire2006a}).
\label{fig:timeseries}}

\end{figure}

\section{Method}

In \cite{Scholling2013} we represented the recruitment process as a walk on a network of recruitment states.
Each cell traverses this network, one state after another. The waiting times $\tau$ between
the state transitions from $i$ to $j$ are considered as a probabilistic event
drawn from an exponential distribution
\begin{equation}
  \rho_{ji}(\tau) = a_{ji} \exp(-a_{ji} \tau),
\end{equation}
where the $a_{ji}$ determine the transition rates. In particular if no transition 
$i \rightarrow j$ is possible, then $a_{ji}=0$. In the case of sequential
recruitment each state has only one possible successor state (compare 
Fig.~\ref{fig:seqrec}).
We showed that the probability of a cell to be in state $j$ is then given by 
Eq.~(\ref{prob}). Since the number of cells in a ChIP experiment is large, $p_k(t)$ does not only represent the 
probability for a single cell to be in state $k$ but also estimates the probability
of occupation of a state within the whole cell population at a given time.

The information of whether a protein $m$ is present in a state $k$ is
represented by a matrix $S_{mk}$:
If the protein is bound in state $k$ then $S_{mk}=1$ and $S_{mk}=0$ otherwise.
By calculating the probability $p_k(t)$ of a cell to be in state $k$ at time $t$,
the resulting ChIP time-series $c_m(t)$ can be predicted by \begin{equation}
c_m(t) = \sum_{k=1}^N S_{mk} p_k(t)\,.
\end{equation}
For a large number of recruitment states ($N \rightarrow \infty$) one
obtains that the ChIP signal for a sequential recruitment process is a 
convolution of the protein binding pattern $S_m(x)$ with the initial
distribution of states $P_0(x)$, where is constant over time ($P_0(x)\equiv p_{xN}(t=0)$,
 see Eq. (\ref{convolution})).
Thus the ChIP signal does not directly represent the binding pattern but is
``blurred'' by $P_0(x)$.

\begin{figure*}[!htpb]
\begin{center}
\includegraphics[width=178mm,height=8cm,keepaspectratio]{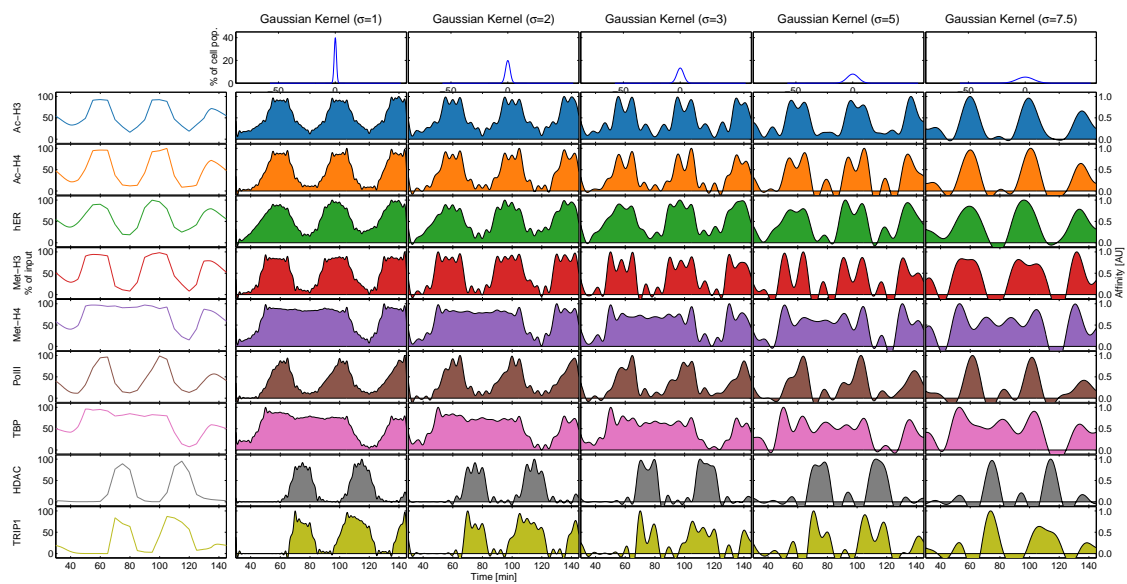}
\end{center}

\caption{Results for Wiener deconvolution (5\% signal-to-noise ratio, 1\textsuperscript{st}
row: deconvolution kernel). 1\textsuperscript{st} column: ChIP time-series
for five proteins and modifications (methylation and acetylation)
of histones \textit{H3} and \textit{H4}. 2\textsuperscript{nd} -- 6\textsuperscript{th} column:
protein binding pattern (normalized) for the deconvolution in Wiener
space with Gaussian kernel with standard deviations $\sigma=\{1\,\textrm{min}, 
2\,\textrm{min}, 3\,\textrm{min}, 5\,\textrm{min}, 7.5\,\textrm{min}\}$. \label{fig:wiener}}
\end{figure*}

$P_0(x)$ can be identified as the convolution kernel, the $S_m(x)$ as the deconvoluted signal and the $c_m(t)$'s as the convoluted signals.
One can therefore apply deconvolution methods to recover the protein binding pattern from the ChIP signal.

Deconvolution methods are frequently used in image processing where the task is
to reconstruct the original image from a blurred image. From a mathematical
point of view, the task is to recover the unknown, un-blurred function
$g(x) \equiv S_m(t)$ from a known, blurred function $f(x)\equiv c_m(t)$ where
$h(x)$ describes the blurring process.
The convolution kernel $h(x)$
is unknown in many applications. To overcome this dilemma, one either
has to assume a certain kernel function or use methods that are able
to estimate the kernel $h(x)$ from the blurred image with the help of certain
assumptions. The latter methods are called \textit{blind deconvolution}
methods \citep{Fish1995,Levin2009}.

In this paper we assume that the kernel is given by a Gaussian distribution,
meaning that some cells are already ahead
in the recruitment process whereas others lag behind.
Based on the results of \cite{Metivier2003} and \cite{Scholling2013},
we assume that the recruitment process is sequential
and the kernel is constant in time, i.e. the de-synchronization happened
prior to the considered time interval that is used for deconvolution.
The latter can be justified by the fact that no obvious decay of signals is detectable
in the ChIP time-series we consider \citep{Scholling2013}.
Thus, no further \mbox{(de-)}synchronization of the cell population
occurs during this time interval.

Note that the analysis of recruitment processes that are non-sequential,
i.e. these processes are not traversed in a cyclic order but contain ``shortcuts''
or alternative recruitment pathways, is beyond the scope of this paper.

An important property of ChIP assays is their sensitivity: the unit
of $c_m(t)$ is expressed in ``percentage of input'', meaning the relative
abundance of a specific DNA sequence using a protein-specific antibody
compared to a non-specific antibody control. This approach leaves
one degree of freedom in amplitude of the ChIP time-series
since different antibodies can have different affinities and thus
cause ChIP signals that differ in amplitude. Consequently, this property
also transfers to the amplitude of the protein binding pattern $S_m(t)$.
For this reason, we normalize the amplitude of ChIP time-series $c_m(t)$ and 
the protein binding pattern $S_m(t)$.

\begin{figure*}[!hbtp]
\begin{center}
\includegraphics[width=178mm,height=8cm,keepaspectratio]{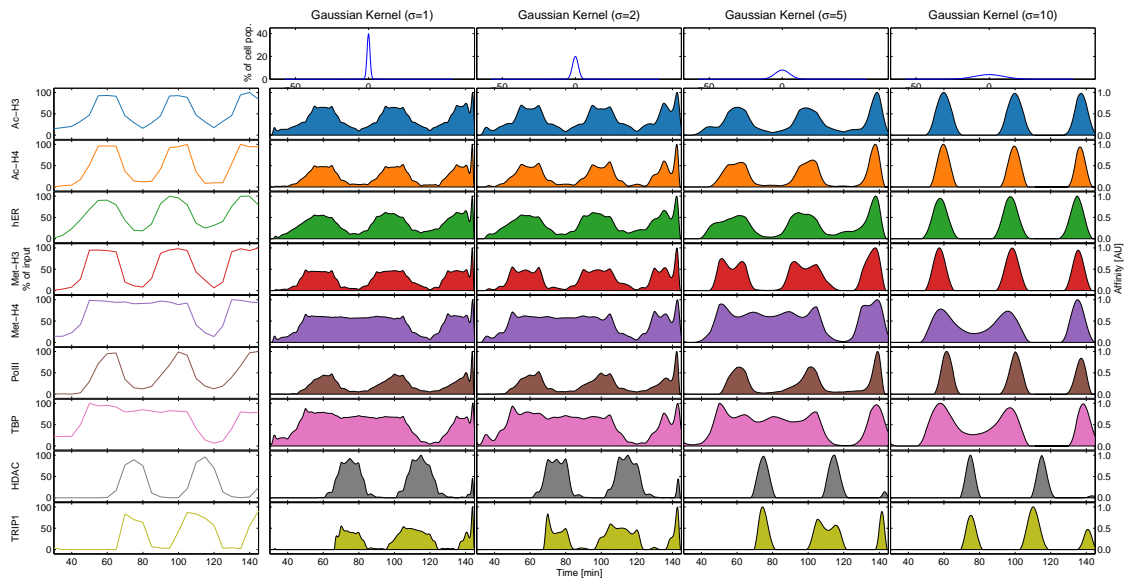}
\end{center}

\caption{Results for Lucy-Richardson deconvolution (same setup as
in Fig.~(\ref{fig:wiener})): For Gaussian kernels with small standard
deviation the binding patterns do not differ strongly from the corresponding
ChIP time-series. The protein binding patterns for larger Gaussian
kernels show distinct peaks for the binding events. The protein binding
patterns for \textit{TBP} and \textit{Met-H4} show plateaus.\label{fig:lr}}
\end{figure*}

In this paper we consider four deconvolution methods:

\begin{itemize}
\item Wiener deconvolution \citep{Dhawan1985} is a deconvolution method that works in
the frequency domain. In contrast to naive deconvolution in Fourier space, it
suppresses amplification of errors at frequencies with a low signal-to-noise ratio.

\item The Lucy-Richardson (LR) deconvolution is a method based on Bayes' theorem.
It maximizes the likelihood of the restored signal to be the true one by using
the expectation-maximization algorithm \citep{Lucy1974}.

\item Blind Lucy-Richardson deconvolution tries to recover the convolution kernel by exploiting the
commutative property of the convolution operation: With an initial guess of the 
kernel, the LR scheme is applied iteratively first to calculate a preliminary
deconvolution of the convoluted signal and then to calculate a new kernel
from this preliminary deconvolution result \citep{Biggs1997}.

\item Binary blind deconvolution \citep{Lam2007} applies a binary constraint on the
deconvoluted signal ($g(x) \in \{0,1\}$). The deconvolution can then be cast
into a convex optimization problem, which can be solved by standard methods for convex
optimization.

\end{itemize}

In contrast to the first two deconvolution methods the latter methods are blind deconvolution methods:
These algorithms try to predict a convolution kernel $h(x)$ instead of using a given kernel as input.
The problem of blind deconvolution is ill-posed in general \citep{Levin2009}:
Many pairs $\{h(x),f(x)\}$ of kernel and un-blurred function exists
that can reproduce the observed, blurred signal $f(x)$. Thus additional
assumptions on the kernel and/or on the un-blurred signal
are necessary.
A detailed description of the deconvolution methods can be found in the
Supporting Information.

We investigate the suitability of these four deconvolution methods
to recover the protein binding pattern from ChIP signals for the \textit{pS2}
gene in the next section.

\begin{figure*}[!htpb]
\begin{center}
\includegraphics[width=178mm,height=8cm,keepaspectratio]{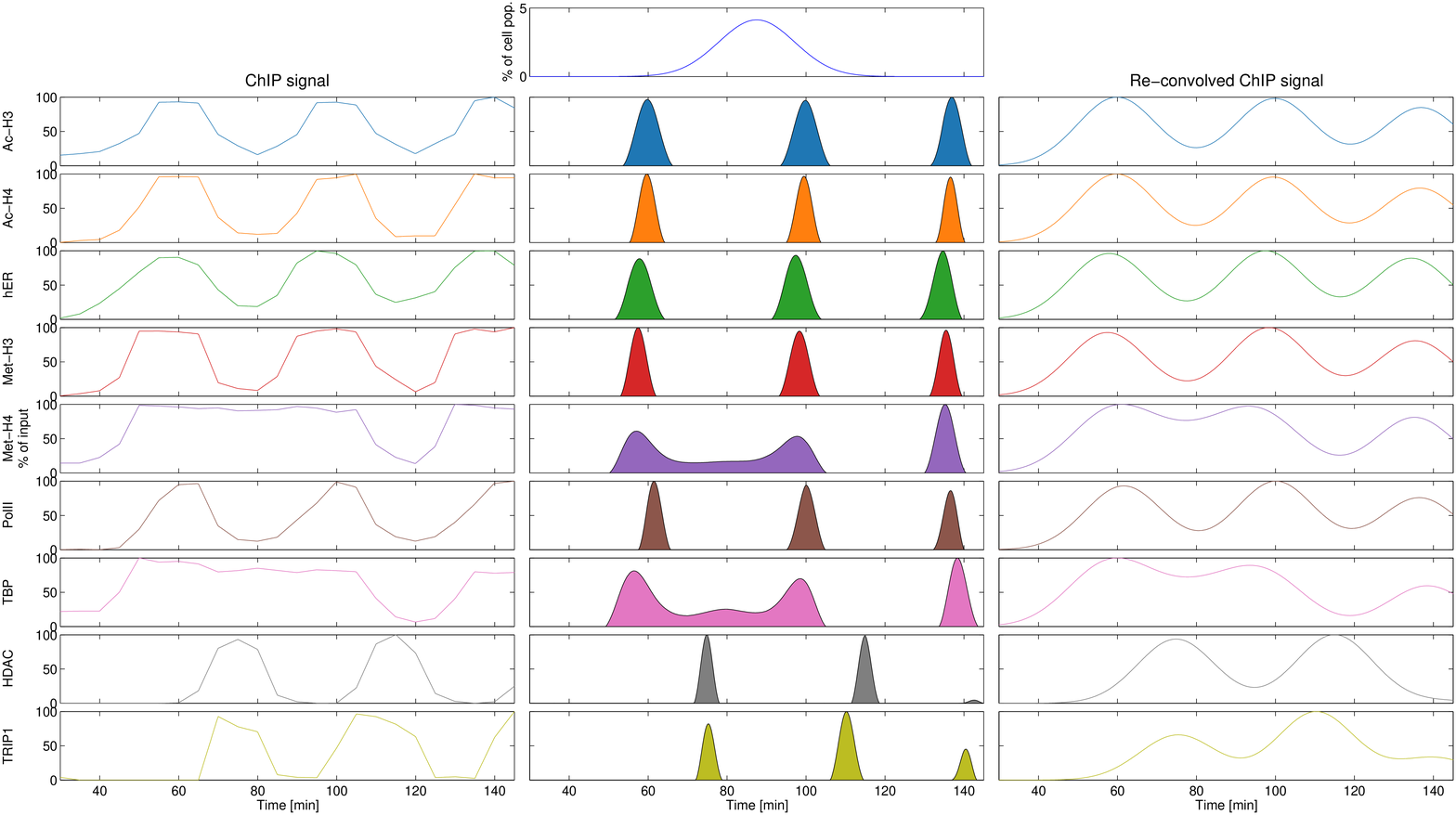}
\end{center}

\caption{Results for Lucy-Richardson deconvolution with the Gaussian
kernel $\sigma=9.69\textrm{min}$) (
1\textsuperscript{st} row: deconvolution kernel). 1\textsuperscript{st} column: ChIP time-series
for five proteins and modifications (methylation and acetylation)
of histones \textit{H3} and \textit{H4}. 2\textsuperscript{nd}--6\textsuperscript{th} column:
protein binding pattern (normalized) for the Lucy-Richardson deconvolution with Gaussian kernel
with standard deviations $\sigma=9.69\,\textrm{min}$. 3\textsuperscript{rd} row: 
result of re-convolving the deconvolved signal with the Gaussian kernel (compare
to 1\textsuperscript{st} column). \label{fig:lr-best}}
\end{figure*}

\section{Results}

We apply the four deconvolution methods to ChIP time-series for the
promoter of the \textit{pS2} gene in human MCF-7 breast cancer cells
\citep{Metivier2003}. Transcription of the \textit{pS2} gene is induced
by the human estrogen receptor \textit{hER}$\alpha$. Here we consider
the binding of five proteins (\textit{hER}$\alpha$\textit{, HDAC,
PolII, TBP}, \textit{TRIP1}) and the histone modifications (acetylation
and methylation) of two histones (\textit{H3} and \textit{H4}). The
time-series of the ChIP data in Fig.~(\ref{fig:timeseries}) covers
a time interval of 145 minutes with an experimental temporal resolution
of approximately 5 minutes and exhibits oscillatory dynamics. We resampled
the data to a time axis with $\Delta t=15\textrm{sec}$ to account
for non-equidistributed time points. Since the first transcription
cycle is unproductive, we only consider the time span 35--145min of
the experiment. A striking feature of the ChIP signals \hilight{in the 
dataset extracted from} \cite{Metivier2003} is that two
of the signals (\textit{Met-H4} and \textit{TBP}) show distinct dynamics
from the other five signals: Instead of three oscillations, only two
oscillations can be observed. Note that exclusion of these two signals
from our analysis does not alter the deconvolution results of the other
signals.

%
%
%

\subsection{Wiener Deconvolution} \hspace{0.1cm}

The results for Wiener deconvolution are shown in Fig.~(\ref{fig:wiener}) for 
Gaussian kernels with standard deviations $\sigma \in \left\{1,2,3,5,7.5\right\}$\,min.
We estimate the signal-to-noise ratio to be 5\%. A signal-to-noise ratio of
10\% only alters the deconvolution results slightly (not shown here).
For kernels with low standard deviation ($\sigma \le 2$\,min) the deconvolution
results barely differ from the ChIP time-series. For larger standard
deviations the deconvolution results in protein binding patterns 
exhibit negative amplitudes. This violates the constraint that concentrations 
can only be non-negative and therefore have to be discarded. 
The results of Wiener deconvolution show that this deconvolution method
\hilight{only results in valid binding patterns for Gaussian kernels with small
standard deviation ($\sigma \le 2 \,$min)} and these binding patterns
show only marginal differences compared to the original ChIP signals.

\subsection{Lucy-Richardson Deconvolution} \hspace{0.1cm}

The results of the Lucy-Richardson deconvolution method are shown
in Fig.~(\ref{fig:lr}). The deconvolved signals for Gaussian
kernels with a small standard deviation ($\sigma<10$\,min) do not
differ strongly from the ChIP signals. Artifacts in the form of large
peaks are visible at the end of the protein binding patterns at $t=145$\,min
for $\sigma\in\{1\,\textrm{min},2\,\textrm{min}\}$. However, for a Gaussian
kernel with large standard deviation ($\sigma=10$\,min) the deconvolution
results in distinct peaks for the protein binding patterns. Inbetween
these two peaks the protein binding patterns decrease to zero (with the
exceptions of \textit{TBP} and \textit{Met-H4} as we discuss below).


Since convolution blurs the signal, we searched for the Gaussian kernel
that results in the sparsest deconvolved signal.
This Gaussian kernel was found at $\sigma=9.69\,\textrm{min}$ and is shown
in Figs. (\ref{fig:lr-best}, \ref{fig:lr-heatmap}).

The signals that show approximately three periods in the ChIP time-series,
also show three distinct binding times in the deconvolution result.
The ChIP signals for methylation of \textit{H4} and binding of \textit{TBP}
show conceptually different dynamics than the ChIP signals of the
other proteins (two oscillations instead of three oscillations). As
a consequence, their binding patterns also show different characteristics
than the binding patterns of the other signals: The first and the
second binding events are not very distinct in time. Inbetween these
two peaks the binding pattern exhibits plateaus in the affinity of about 0.2
and 0.15 (arbitrary units) for \textit{TBP} and \textit{Met-H4}, respectively.

\subsection{Blind Lucy-Richardson Deconvolution} \hspace{0.1cm}

\begin{figure*}[!htpb]
\begin{center}
\includegraphics[width=178mm,height=8cm,keepaspectratio]{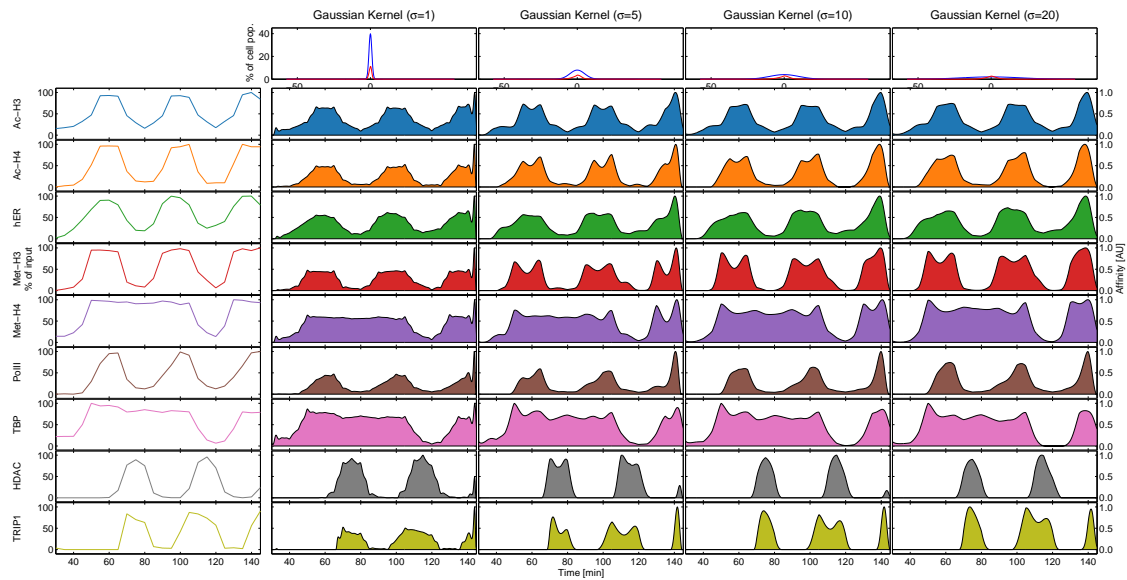}
\end{center}

\caption{Results for blind Lucy-Richardson deconvolution (same setup as
in Fig. \ref{fig:wiener}): 1\textsuperscript{st} row: The initial kernels
are plotted in blue and the final estimation of the kernels are plotted
in red. For all kernels the protein binding patterns show strong similarities
to the original ChIP time-series.\label{fig:lr-blind}}
\end{figure*}

The results for the blind Lucy-Richardson deconvolution are shown
in Fig. (\ref{fig:lr-blind}). Note that blind Lucy-Richardson deconvolution
starts from an initial guess of the kernel and estimates the un-blurred
signal and the convolution kernel with the maximal likelihood. We
started the deconvolution with Gaussian kernels. The resulting kernels
are shown in the first row of Fig. (\ref{fig:lr-blind}) indicated
in red. The deconvolved signals are very similar to the original ChIP signal
for all kernels. Interestingly, for initial Gaussian kernels with
large standard deviation ($\sigma\ge15\,\textrm{min}$) the initially broad kernels are 
always driven towards sharply peaked distributions, which explains the small effect 
yielded by this method.

\subsection{Binary Blind Deconvolution} \hspace{0.1cm}

The deconvolution algorithm proposed by \cite{Lam2007} performed worst:
The algorithm failed to converge for the ChIP time-series for initial
Gaussian kernels. For this reason, we can not present any protein
binding patterns for this deconvolution method.

\section{Discussion}

\hilight{
Three of the four deconvolution algorithms we tested were able to deconvolve the protein binding patterns from the ChIP signals.
However, Wiener deconvolution only producted valid binding patterns for sharp convolution kernels (small $\sigma$) and also blind Lucy-Richardson deconvolution resulted in sharp convolution kernels.
}

As one may expect \hilight{deconvolution with these kernels does} not show substantial differences to the original ChIP signals.
This can be explained by the fact that these Gaussian kernels are very
near to the Dirac delta function $\delta(x)$. Note that convolution
of a signal with the Dirac delta function results in the very same
signal again ($\left(f\otimes\delta\right)(x)=f(x)$).

\hilight{Non-blind Lucy Richardson deconvolution resulted in protein binding 
patterns that did substantially differ from the ChIP signals.}
Regarding the results for the kernel that reproduces the sparsest admissible protein
binding pattern ($\sigma=9.69$\,min), the first events that occur in the recruitment
process are binding of \textit{TBP} and methylation of histone \textit{H4}.
However, these events are not indicated very sharply in the deconvolved
signal and thus their binding times are less reliable. We defer the
discussion of the deconvolution results for these events to the end
of this section.

\begin{figure}[!bt]
\includegraphics[width=86mm]{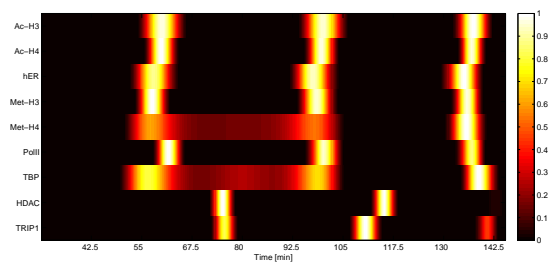}

\caption{Results for Lucy-Richardson deconvolution with the Gaussian
kernel $\sigma=9.69\,\textrm{min}$ (shows the same data as 2\textsuperscript{nd} column of Fig. \ref{fig:lr-best}): blue areas indicate low binding affinity, red areas indicate
high binding affinities.\label{fig:lr-heatmap}}
\end{figure}

The protein binding patterns for the other four binding and histone
modification events are indicated more distinctly than for \textit{Met-H4}
and \textit{TBP}. According to the deconvolutions in Fig. (\ref{fig:lr-heatmap}),
the event that occurs first is the binding of \textit{hER}$\alpha$%
\footnote{For the first productive cycle one can question whether the binding
of \textit{hER}$\alpha$ or the methylation of histone \textit{H4}
occurs first. But in the second and third cycle a clear distinction
is possible.%
}. This result is consistent with prior biological knowledge, since
the transcription of \textit{pS2} is known to be initiated by the binding
of this estrogen receptor. Next in the recruitment process the histone
\textit{H3} is methylized, which is indicated by sharp peaks in each
of the three recruitment cycles. \textit{CBP} is known to be capable
of methylizing this histone \citep{Wang2001}. However, \textit{CBP}
is not included in the present experimental data and thus we cannot check
whether \textit{CBP} is accountable for this methylation event.

Deacetylation events of \textit{H3} and \textit{H4} follow according
to the results of LR deconvolution. These events end just when the
binding pattern of RNA polymerase II peaks. Interestingly, \citet{Metivier2006}
defines a transcriptionally engaged \textit{pS2} promoter by the presence
of \textit{Met-H4} and \textit{Ac-H3}. In the results of LR deconvolution
the methylation time of \textit{H4} is not indicated very precisely,
but the time of deacetylation for \textit{H3} matches with the binding
of RNA polymerase II.

The re-convoluted signals of \textit{HDAC} and \textit{TRIP1} exhibit the largest
errors in reconstructing the initial ChIP time-series (Fig. \ref{fig:lr-best}) \hilight{and regarding to our results they bind very late in the recruitment process.
It is known that these proteins are involved in promoter clearance 
}\citep{Metivier2003}\hilight{ which is in agreement with the late binding time predicted by our results.
However, the predicted binding time of \textit{HDAC} (histone deacetylase) 
does not coincide with the deacetylation of \textit{H3} and \textit{H4}.

Thus we can propose two hypotheses:
Either this means that \textit{HDAC} does not require to bind to the protein
complex to deacetylize the histones or \textit{HDAC} is bound while histone deacetylation but the Lucy-Richardson prediction of the beginning of
the \textit{HDAC} binding time is incorrect.

In this case the binding of \textit{HDAC} better coincides with the deacetylation events for
kernels with small standard deviation $(\sigma \le 2\,$min) e.g. as predicted by blind Lucy-Richardson.
Thus the second possibility would indicate that cell sample used in the ChIP experiment is strongly synchronized.
}


We return to the discussion of \textit{Met-H4} and \textit{TBP}, which
show distinct dynamics in the ChIP signals: instead of three oscillations
like the other time-series they only show two oscillations. \citet{Metivier2003}
argued that detachment of \textit{TBP} and demethylation of
\textit{H4} do not occur in each productive recruitment cycle. Instead
these events occur only in every second cycle. Besides the strict
alternation of even and odd cycles, also probabilistic
alternation is a possible interpretation of these binding patterns
\citep{Scholling2013}. In this context, one
can interpret the plateaus between the two peeks for \textit{TBP}
and \textit{Met-H4} in Figs. (\ref{fig:lr}, \ref{fig:lr-heatmap})
as signals only from those cells in which \textit{TBP} does not detach
and \textit{H4} is not de-methylized after the first productive cycle.

\section{Conclusions}

In the present paper, we have analyzed the applicability of different
deconvolution methods to recover the protein binding pattern from
ChIP time-series with an imperfectly synchronized cell population.
Our analysis showed that Lucy-Richardson deconvolution can be used
to improve inference \hilight{on} protein binding pattern from ChIP
time-series. The presented results \hilight{allow} a more \hilight{precise} 
interpretation of the ChIP data \hilight{in forms of the function proteins 
play in the recruiting
protein complex and the recruitment process in general.} We considered
two non-blind deconvolution methods, Wiener deconvolution
and Lucy-Richardson deconvolution, as well as two blind deconvolution
methods, blind Lucy-Richardson deconvolution and binary blind deconvolution.
The applicability of these methods was tested on ChIP data for the
\textit{pS2} gene. \hilight{Binary blind deconvolution failed
to perform deconvolution. Wiener deconvolution and blind Lucy-Richardson 
deconvolution resulted in protein binding patterns that did not differ 
substantially from the original ChIP signals. For Wiener deconvolution broader
convolution kernels violate the positivity constraint of the deconvoluted signal.
However, the Lucy-Richardson deconvolution method
has proofed its potential to recover protein binding patterns 
with distinct binding times.
The binding pattern of \textit{HDAC} indicates a late binding time
which corresponds to its
clearance function in the recruitment process but does not coincide with the
deacetylation events of \textit{H3} and \textit{H4}.
This can either be interpreted as an inaccurate prediction of the proteins'
binding pattern for this deconvolution method or it means that \textit{HDAC}
does not require to bind to the promoter to deacetylize the histones.
}
The reconstruction of the binding patterns succeeded for all other
proteins. Their timing is in agreement with prior biological knowledge and can
help to investigate the exact function of the proteins in the recruitment process.

\paragraph{Funding\textcolon}

This work has been supported by the \textit{Forum Integrativ\-medizin}
an initiative of the \textit{Hilde Umdasch Privatstiftung}.

\bibliographystyle{natbib}
\bibliography{library}

\end{document}